\documentclass[aps,prl,reprint,groupedaddress,superscriptaddress, amsmath, amssymb, longbibliography]{revtex4-2}

\usepackage{graphicx}
\usepackage{dcolumn}
\usepackage{bm}
\usepackage{hyphenat}

\begin{document}


\title{Quasi-static strain governing ultrafast spin dynamics}




\author{Y. Shin}
\affiliation{Department of Physics, Kunsan National University, Kunsan 54150, South Korea}
\affiliation{Department of Physics, Chungbuk National University, Cheongju 28644, South Korea}

\author{M. Vomir}
\affiliation{Université de Strasbourg, CNRS, Institut de Physique et Chimie des Matériaux de Strasbourg,\\ UMR 7504, Strasbourg 67034, France}

\author{D.-H. Kim}
\affiliation{Department of Physics, Chungbuk National University, Cheongju 28644, South Korea}

\author{P. C. Van}
\affiliation{Department of Materials Science and Engineering, Chungnam National University, Daejeon 34134, South Korea}

\author{J.-R. Jeong}
\affiliation{Department of Materials Science and Engineering, Chungnam National University, Daejeon 34134, South Korea}

\author{J.-W. Kim}
\email[Corresponding author$:$ ]{hwoarang@kunsan.ac.kr}
\affiliation{Department of Physics, Kunsan National University, Kunsan 54150, South Korea}



\date{\today}

\begin{abstract} 
The quasi-static strain (QSS) is the product generated by the lattice thermal expansion after ultrafast photo-excitation and the effects of thermal and QSS are inextricable. Nevertheless, the two phenomena with the same relaxation timescale should be treated separately because of their different fundamental actions to the ultrafast spin dynamics. By employing ultrafast Sagnac interferometry and magneto-optical Kerr effect, we quantitatively prove the existence of QSS, which has been disregarded, and decouple two effects counter-acting each other. Through the magnetoelastic energy analysis, rather we show that QSS in ferromagnets plays a governing role on ultrafast spin dynamics, which is opposite to what have been known on the basis of thermal effect. Our demonstration provides an essential way of analysis on ultrafast photo-induced phenomena.
\end{abstract}


\maketitle


Magnetoelasticity, the coupling of the spin and the strain, is a universal phenomenon in magnetic materials and enables an active control of spin states by modifying material dimensions \cite{Kittel1958}. Until now, it has been described that the spin dynamics, after ultrafast spin angular momentum transfer by photo-excitation \cite{Beaurepaire1996, Koopmans2000, Zhang2000, Kirilyuk2010, Hofherr2020, Gort2018}, is governed by time-dependent effective fields with origins such as magnetocrystalline \cite{Bigot2005}, dipole, Zeeman \cite{Hohlfeld2001}, exchange \cite{Radu2011, Mathias2012, Batignani2015}, terahertz \cite{Kampfrath2011, Baierl2016}, magnetoelasticity \cite{Scherbakov2010, Kim2012}, spin current \cite{Nvemec2012, Huisman2016, Choi2020} etc. These have been mainly described via electron and spin degrees of freedoms and while, due to comparatively slower response, the lattice degree of freedom is being highlighted only recently \cite{Dornes2019}.\\
\indent The increase of the lattice temperature by photo-excitation generates two types of strain, which are the propagating one with a temporal width of a few ps and the QSS induced from long-lived thermal expansion of lattice. In the recent decade, the interaction of spins with the propagating strain has been intensively investigated from diverse points of view \cite{Kovalenko2013, Januvsonis2016, Thevenard2016}, however, QSS has never been regarded despite its comparable amplitude to that of the propagating one. Its principal reason is that since QSS and thermal energy are inextricable as thermal expansion produces QSS at all times, an effect of QSS has been misled to be a well-known temperature effect. As a simple example, we have used the phenomenological three-temperatures model to extract the time-dependent temperatures information of the sub-systems through fitting with the experimental curves of spin and reflectivity dynamics. However, the model does not contain the information of strains that exist as long as the thermal expansion does. This fact might deliver improper messages and hinder unveiling new physics of ultrafast dynamics. As two effects have different contributions to dynamics (the mechanical pressure of the former changes plasmonic \cite{Kim2016} or electronic bands \cite{Akimov2006} and the latter does electron populations), systematic and comprehensive measurement approaches to separate two effects are required for the complete analysis of the fundamental mechanism of the ultrafast photo-induced spin dynamics.\\
\indent In this Letter, using ultrafast Sagnac interferometry (USI) as well as magneto-optical Kerr effect (MOKE), we unambiguously demonstrate that the QSS in ferromagnetic films governs the ultrafast spin dynamics from the first ps to the sub-ns timescale. In particular, the Sagnac interferometry was used to directly obtain the lattice expansion dynamics. Contrary to the known beliefs, we experimentally revealed three important evidences as follows: i) the increase of the spin precession frequency with the pump intensity, ii) $\pi$-phase inversion of the precession, iii) the pronounced background distinguished from incoherent magnons. The model calculation of Landau-Lifshitz-Gilbert (LLG) equation incorporating the time-dependent QSS strongly supports that all features mentioned are elucidated by one concept of QSS. With ferromagnets (Co, Ni, Ni$\textsubscript{x}$Fe$\textsubscript{1-x}$) to vary the competition between strain and thermal effects, our scenario is consolidated with full consistency.
\begin{figure}
\includegraphics[width=0.45\textwidth]{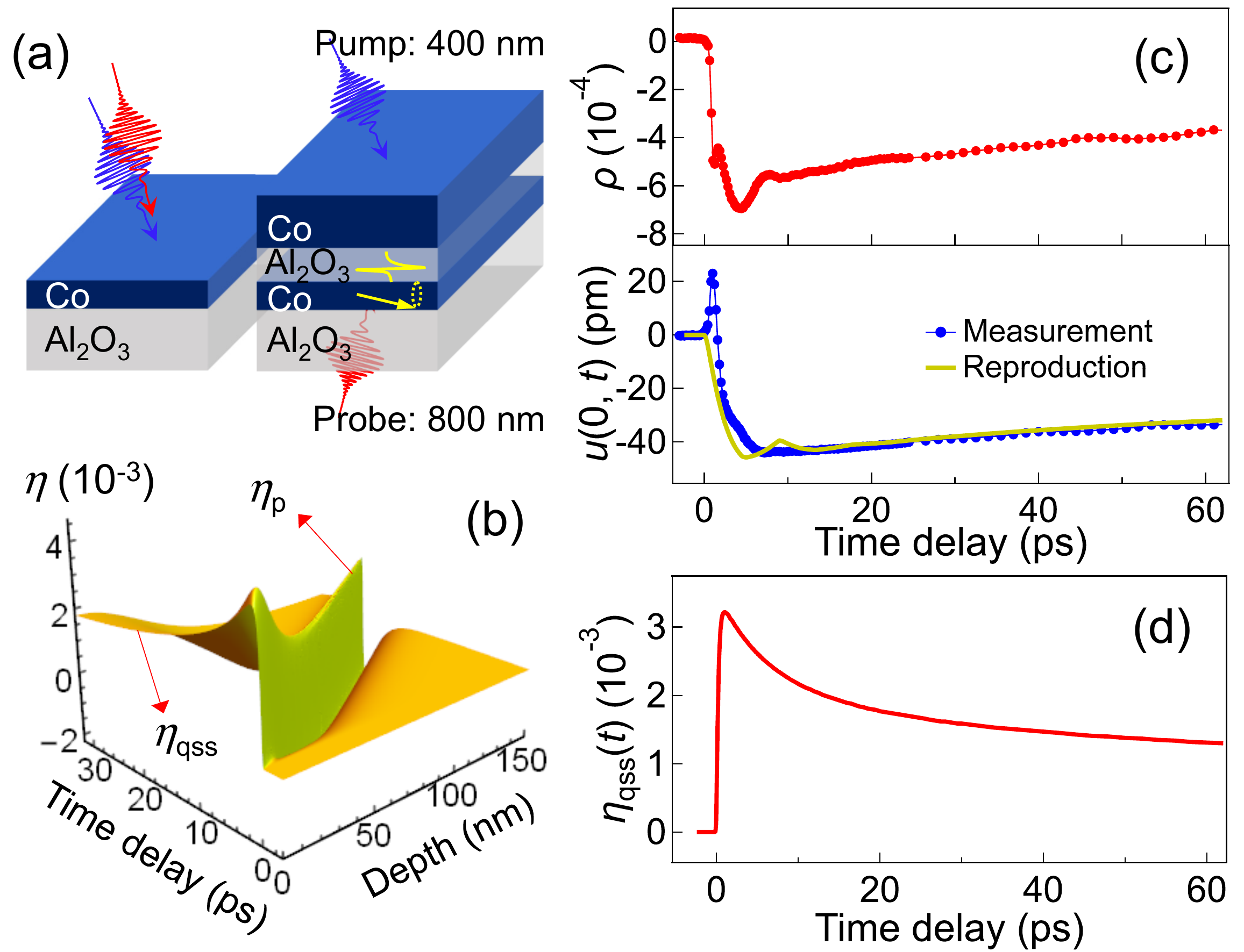}
\caption{Measurement geometry of the experiment and ultrafast Sagnac interferometric data. (a) Simple pictures of the pump-probe MOKE measurements: the pump and probe beams hit the top surface for a single layer (left) and a pump excites the top surface and a probe beam measures the dynamics of the back side for a trilayer (right).  (b) Dynamic strain profile $\eta(z, t)$ calculated for a thick Co film: the quasi-static strain $\eta\textsubscript{qss}(z, t)$ and the propagating bipolar pulse $\eta\textsubscript{p}(z, t)$ are shown. (c) Sagnac interferometric curves: real ($\rho$: upper graph) and imaginary parts ($\delta\phi$: bottom graph) of the complex reflectivity change $\delta$$r/r$. The lattice displacements $u(0, t)$ converted from experimental data $\delta\phi$ (blue circles) and reproduced by the calculation (yellow line) are displayed for the left axis in the bottom graph. (d) $\eta\textsubscript{qss}(t)$ profile extracted from $u(0, t)$. The tensile strain (positive sign) stands for the lattice expansion by temperature increase.}\label{fig:fig1}
\end{figure}\\
\indent The measurement geometry of pump-probe MOKE is shown in Fig. 1(a). We prepared the single layer of Al$_{2}$O$_{3}$(5 nm)/Co(25 nm)/Al$_{2}$O$_{3}$ and the trilayer of Co(200 nm)/Al$_{2}$O$_{3}$(15 nm)/Co(25 nm)/Al$_{2}$O$_{3}$ by magnetron sputtering (nm in thickness are omitted hereafter). The Al$_{2}$O$_{3}$ in the single layer was deposited to protect from oxidation and that in the trilayer are used for the acoustic impedance matching with Co and the suppression of possible propagation of thermal magnons from the top Co layer. We used Ti:sapphire regenerative amplified laser pulses with a repetition rate of 10 kHz, a temporal width of 40 fs, and a center wavelength of 800 nm. The frequency-doubled pump pulses ($\lambda\textsubscript{pu}$ = 400 nm) excite front sides and probes ($\lambda\textsubscript{pr}$ = 800 nm) measure the differential Kerr rotation $\Delta$$\theta$$(t)$ and reflectivity $\Delta$$R(t)$ either of the front side in the single layer or the back side in the trilayer. Figure 1(b) presents the simulated strain profile $\eta$$(z, t)$ for a thick Co slab, showing both a bipolar strain pulse $\eta$$\textsubscript{p}(z, t)$ propagating into the depth and a quasi-static strain $\eta$$\textsubscript{qss}(t)$ (= $\eta$$(0, t)$) near a sample surface with a long decay time of thermal energy.\\
\indent The main purpose of ultrafast Sagnac interferometry \cite{Tachizaki2006}, based on an oscillator system, is to prove the existence of the QSS by quantitatively measuring the lattice displacement dynamics $u(z = 0, t)$ and finally extract $\eta$\textsubscript{qss}$(t)$ (see the Supplemental Material SM1 for characteristics of the instrument \cite{Supp}). In general, it is not obvious to extract the lattice temperature profile from $\Delta$$R(t)$, which will be the driving source for $u(z = 0, t)$ in one-dimensional wave equation, due to unknown piezo-optic property of materials except Ni, Cr, and Au \cite{Saito2003, Pezeril2014}. Therefore, we first measured the demagnetization in a longitudinal geometry $\Delta$$\theta$$\textsubscript{L}(t)$/$\theta$$\textsubscript{L}$ indicating the spin temperature under in-plane field of $H$$\textsubscript{ext}$ = 0.5 T with three-temperatures model. The pump intensity calibration for USI and MOKE was performed by comparing $\Delta$$R$/$R$ measured by respective instruments. This allows us to get the good values of both lattice temperature profile and dynamic heat coupling coefficients. Then, after matching the solution $u(z = 0, t)$ of one-dimensional wave equation with experimental data from ultrafast Sagnac interferometry, $\eta$$\textsubscript{qss}$ is unambiguously determined (see the Supplemental Material SM2 for calculation details \cite{Supp}). Figure 1(c) shows the real part ($\rho$ = $\Delta$$R$/$2R$: upper panel) which is related to the differential reflectivity and the converted lattice displacement ($u(0, t)$ = $\lambda$$\textsubscript{pr}$$\delta\phi$/4$\pi$) from the imaginary part ($\delta\phi$) of the complex reflectivity change $\delta$$r$/$r$ $\sim$ $\rho$ + $i$$\delta\phi$ (blue circles for measurement and the yellow line for reproduction from the modelling). Here, $\rho$ is defined as the relative change in reflectance, and $\delta\phi$ the phase change induced by the pump.
\begin{figure}
\includegraphics[width=0.42\textwidth]{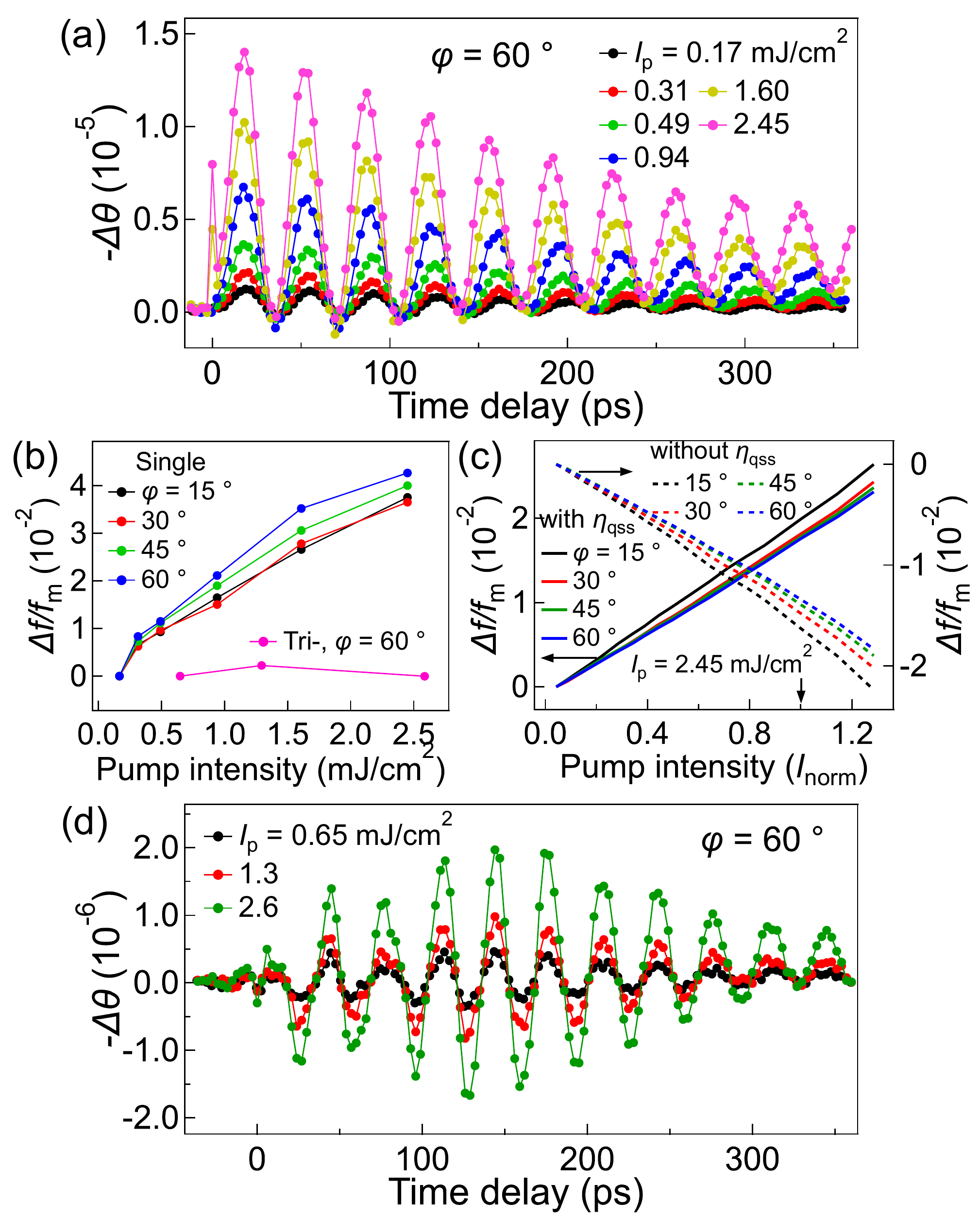}
\caption{Spin precession frequency as a function of pump intensity of a Co layer. (a) Differential magneto-optical Kerr rotations $\Delta\theta(t)$ of a Co(25) single layer as a function of $I$$\textsubscript{p}$ under $H$$\textsubscript{ext}$ = 0.4 T and $\varphi$ = 60 $^{\circ}$. (b) Summary of the frequency change of the spin precession $\Delta$$f/f$$\textsubscript{m}$ for a single layer as a function of $I$$\textsubscript{p}$ and $\varphi$ (15 $^{\circ}$: black, 30 $^{\circ}$: red, 45 $^{\circ}$: green, 60 $^{\circ}$: blue circles). The pink circles present $\Delta$$f/f$$\textsubscript{m}$ of Co(25) in a trilayer. (c) Model calculations of $\Delta$$f/f$$\textsubscript{m}$ based on LLG equation in the presence (solid lines) and absence (dashed lines) of QSS, respectively. $I$$\textsubscript{norm}$= 1.0 corresponds to $I\textsubscript{p}=2.45$ mJ/cm$^{2}$ used in the experiment. (d) Differential Kerr rotation $\Delta\theta(t)$ induced by $\eta\textsubscript{p}$ for a trilayer under $H$$\textsubscript{ext}$ = 0.4 T and $\varphi$ = 60 $^{\circ}$. }\label{fig:fig2}
\end{figure}\\
\indent Using the relation of $u(0, t)$=$\int_{z=0}^{d}\eta$$(z', t)$$\mathrm{d}z'$, we extracted $\eta$$\textsubscript{qss}$ as plotted in Fig. 1(d). The positive value of $\eta$$\textsubscript{qss}$ stands for a tensile strain resulted from the lattice expansion by sign convention. We note that the experimental data $u(0, t)$ in Fig. 1(c) shows a sharp positive peak at first ps implying the lattice contraction. This may be explained by several possibilities such as the electronic stress \cite{Wright1994}, the magnetoelastic stress \cite{Reid2018, Von2020}, and the lateral expansion \cite{Von2020, Von2018}, which are actively in discussion. However, identifying the origin is beyond the current scope of our paper and the positive peak has not been considered in the reproduction process.\\
\indent The solid circles in Fig. 2(a) show $\Delta\theta(t)$ of the Co single layer under $H$$\textsubscript{ext}$ = 0.4 T and $\varphi$ = 60 $^{\circ}$ with various pump intensities $I\textsubscript{p}$. The experimental curves for other angles ($\varphi$ = 15, 30, and 45 $^{\circ}$) are shown in the Supplemental Material SM3 \cite{Supp}. As summarized in Fig. 2(b) after a damped sinusoidal function fit, the frequency change of the spin precession $\Delta$$f/f\textsubscript{m}$ increases with $I\textsubscript{p}$ for broad magnetic field angles $\varphi$ (15 $^{\circ}$: black, 30 $^{\circ}$: red, 45 $^{\circ}$: green, 60 $^{\circ}$: blue circles). The each value of $f\textsubscript{m}$ at the minimum $I\textsubscript{p}$ is 13.8, 20.0, 24.4, and 27.6 GHz for the relevant angles of $\varphi$, respectively. This frequency variation observed for the Co film, one of the typical ferromagnets, is opposite to what one would expect by considering the conventional magnetic energy terms of the magneto-crystalline, dipolar, and Zeeman energy. The magnitude of the effective magnetic field $H\textsubscript{eff}(t)$ in such case would decrease with the temperature increase of sub-systems.\\
\indent Let us now focus on the relation between $\eta\textsubscript{qss}(t)$ and $f$. Using LLG equation combined with $\eta\textsubscript{qss}(t)$ and temperature-dependent sub-systems, $\Delta$$f/f\textsubscript{m}$ was calculated. In a magnetic free energy we consider the magnetoelastic energy term $F\textsubscript{me}=-(3/2)\lambda\textsubscript{s}\sigma\textsubscript{s}$cos$^{2}\vartheta$, where $\lambda\textsubscript{s}$ is the magnetostriction, $\sigma\textsubscript{s}=3(1-\nu)B\eta\textsubscript{qss}(t)/(1+\nu)$ the mechanical stress, $\nu$ Poisson’s ratio, $B$ the bulk modulus, and $\vartheta$ the angle between the magnetization $\vec{M}$ and $\sigma\textsubscript{s}$ direction (see the Supplemental Material SM2 for strain calculation and parameter values \cite{Supp}). As the results of the model calculation is plotted in Fig. 2(c), $\Delta$$f/f\textsubscript{m}$ with the inclusion of $\eta\textsubscript{qss}(t)$ (solid lines) are in good agreement with experimental data of Fig. 2(b), showing the increase of $\Delta$$f/f\textsubscript{m}$ up to 2 $\%$ for a wide range of $\varphi$. In contrast, for the absence of $\eta\textsubscript{qss}(t)$, $\Delta$$f/f\textsubscript{m}$ decreases by 1.6 $\%$ (dashed lines) as expected in a conventional analysis. We note that $I\textsubscript{norm}$, the intensity normalized by $I\textsubscript{p}=2.45$ mJ/cm$^{2}$, in the calculation is not arbitrarily chosen but determined by matching the calculated spin temperature with the demagnetization curve $\Delta$$\theta$$\textsubscript{L}(t)$/$\theta$$\textsubscript{L}$.
We checked the case when the spin precession is not involved in neither thermal nor QSS effects. This requirement is met by the back-side measurement of the Co trilayer structure. The strain pulse $\eta$$\textsubscript{p}(z, t)$ generated from the Co top layer arrives at the Co underlayer without carrying thermal energy and contributes only to triggering the spin precession through magnetoelastic coupling. As $\Delta$$\theta$$(t)$ on $I\textsubscript{p}$ under $\varphi$ = 60 $^{\circ}$ is presented in Fig. 2(d) and with the pink curve in Fig. 2(b) up to $I\textsubscript{p}=2.6$ mJ/cm$^{2}$ that corresponds to $\sim$ 70 $\%$ of the burning threshold for the Co top layer, $\Delta$$f/f\textsubscript{m}$ did not show a noticeable change within the experimental uncertainty.
\begin{figure}
\includegraphics[width=0.40\textwidth]{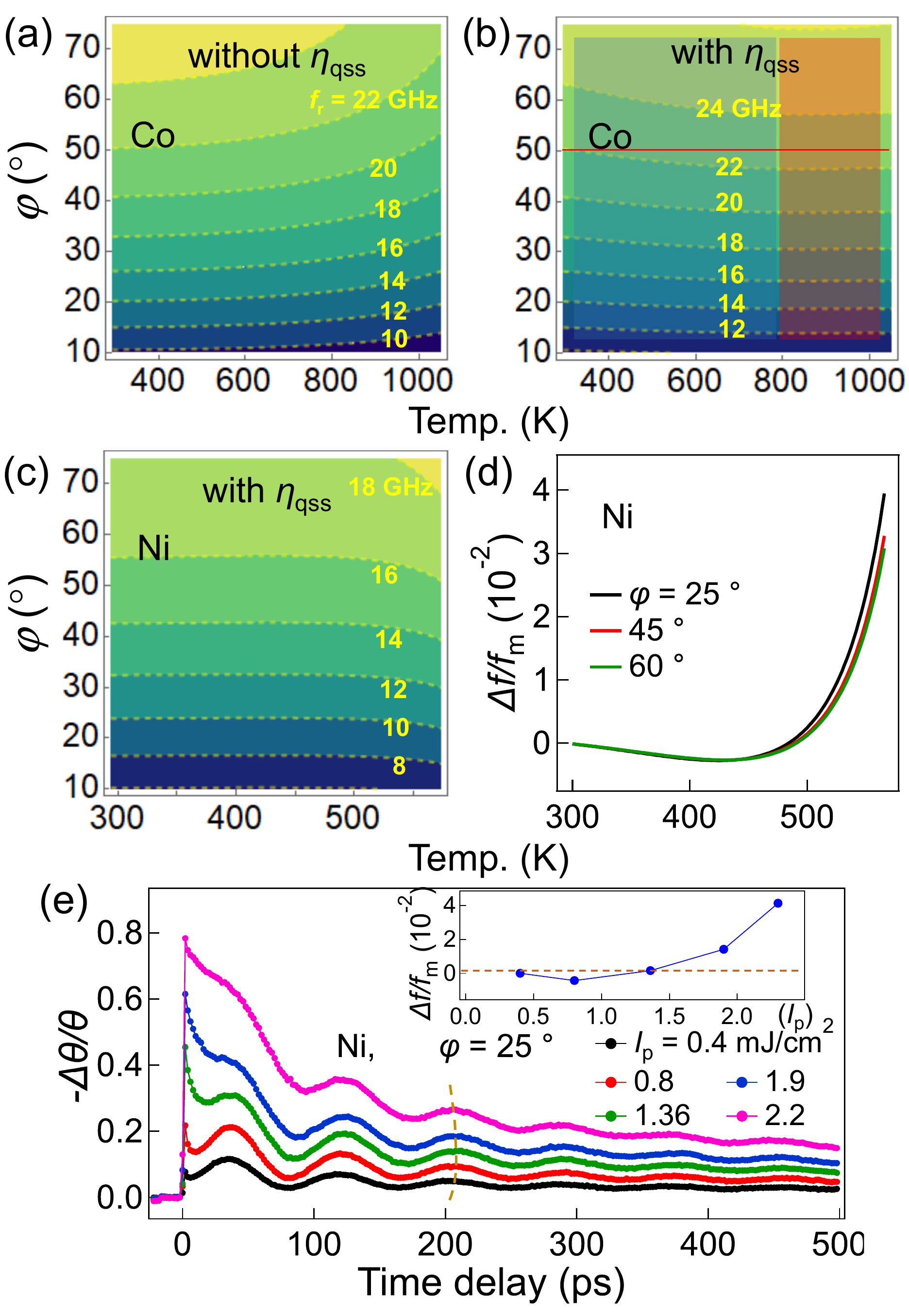}
\caption{Ferromagnetic resonance frequency ($f$$\textsubscript{r}$) variation with control parameters $T$, $\varphi$, and $\eta\textsubscript{qss}$. Contour maps of $f$$\textsubscript{r}$ calculated using Kittel relation as a function of $T$ and $\varphi$ under $H$$\textsubscript{ext}$ = 0.4 T, for the absence ((a) for Co) and presence ((b) for Co and (c) for Ni) of QSS, respectively. The shaded boxes in (b) stand for domains where $f$$\textsubscript{r}$ increases (blue) and starts to slow down (red). (d) Selected curves of $\Delta$$f/f$$\textsubscript{m}$ at $\varphi$ =25, 45, and 60 $^{\circ}$ from Fig. 3(c). (e) $\Delta\theta(t)$/$\theta$ of Ni(270)/SiO$_{2}$ under $H$$\textsubscript{ext}$ = 0.4 T and $\varphi$ = 25 $^{\circ}$ as a function of $I$$\textsubscript{p}$. The sign change of $\Delta$$f/f$$\textsubscript{m}$ appears as shown in the inset.}\label{fig:fig3}
\end{figure}\\
\indent For the comprehensive picture of the QSS effect on $f$ over a wide range of temperatures, we investigate the ferromagnetic resonance frequency ($f\textsubscript{r}$) within the classical Kittel equation $f\textsubscript{r}$ = $\gamma(F_{\theta\theta}F_{\varphi\varphi}-F_{\theta\varphi}^{2})^{1/2}$/(2$\pi$$M(T)$sin$\theta\textsubscript{eq}$) with the same energy terms in the magnetic free energy $F$ as in the dynamics case. The QSS is estimated here by a thermal strain $\eta\textsubscript{th}(T)$ = $\beta$$\Delta$$T$  (thermal expansion coefficient of Co: $\beta$ = 13.7 $\mu$ \cite{Rao1977}) supposing a thermal equilibrium among the sub-systems. Figures 3(a) and 3(b) present the contour map of $f\textsubscript{r}$ with control parameters of $T$ (x-axis) and $\varphi$ (y-axis) under $H\textsubscript{ext}$= 0.4 T in the absence and presence of $\eta\textsubscript{th}(T)$, respectively. In Fig. 3(a), $f\textsubscript{r}$ decreases monotonically as $T$ increases for all values of $\varphi$. On the other hand, Fig. 3(b) clearly indicates that at a fixed value of $\varphi$ (along the red line), $f\textsubscript{r}$ increases gently with the increase of $T$ up to $\sim$ 800 K (blue box region). This is attributed that $\eta\textsubscript{th}(T)$ increases almost linearly with $T$ unlike hardly altered $M(T)$ due to its high Curie temperature ($T\textsubscript{c}$), hence the effect of magnetoelasticity plays a dominant role over those of conventional magnetic energy terms. At high $T$ (orange box region), the rapid drop of $M(T)$ reduces the effective field leading to the inflection point of $f\textsubscript{r}$.\\
\indent It is fairly conjectured that this is not the special case only for Co but can be generalized to a broad class of magnetoelastic materials. We can expect that for low $T\textsubscript{c}$ materials, the faster drop of $M(T)$ shows up at a low $T$, where QSS effect on $f\textsubscript{r}$ is comparatively weaker. The Ni has a low $T\textsubscript{c}$ = 630 K and $M$ = 525 emu/cm$^{3}$ which are key parameters for the thermal effect but similar values of $\lambda\textsubscript{s}$, $\beta$, and $B$ to Co, which are related to the strain effect. The contour map of $f\textsubscript{r}$ including QSS effect for Ni is shown in Fig. 3(c) and $\Delta$$f/f\textsubscript{m}$ for $\varphi$ = 25, 45, and 60 $^{\circ}$ selected from the contour map are plotted in Fig. 3(d). As we see clearly, $f\textsubscript{r}$ decreases first at a low $T$ and is switched to the increase as $T$ increases. To prove those results experimentally, $\Delta\theta(t)$ of Ni(270)/SiO$_{2}$ at $\varphi$ = 25 $^{\circ}$ and $I\textsubscript{p}$ $\le$ 2.2 mJ/cm$^{2}$ was measured and plotted in Fig. 3(e). As presented in the inset of Fig. 3(e), $\Delta$$f/f\textsubscript{m}$ goes to negative and then turns to positive as $I\textsubscript{p}$ increases, meaning that the thermal effect prevails over QSS effect at a low temperature for low $T\textsubscript{c}$ materials. These curves reproduce the calculation (Fig. 3(d)) in excellent agreement supporting our analysis on thermal and strain effects.\\
\indent We address that $\eta$$\textsubscript{qss}$ has a decisive role on the initial stage of spin dynamics by further features that have been overlooked. The first point to remark is the phase of the spin precession. Figure 4(a) displays the calculated curves of $\Delta$$M\textsubscript{z}$$(t)$/$M\textsubscript{z}$ in the absence (red) and presence (blue curve) of $\eta$$\textsubscript{qss}(t)$ under $H\textsubscript{ext}$= 0.54 T and  $\varphi$ = 60 $^{\circ}$. The earlier results of spin precessions in Co and Ni as shown in Fig. 2(a) and 3(e) correspond to the blue curve leading to a $\pi$-phase inversion compared to the conventional free energy analysis. This feature is not attributed to the left-handed precession which recently observed in the specified condition for the sign reversal of the gyromagnetic ratio \cite{Kim2020}, but surely to the opposite reorientation of $H\textsubscript{eff}(t)$ due to the QSS.
\begin{figure}
\includegraphics[width=0.37\textwidth]{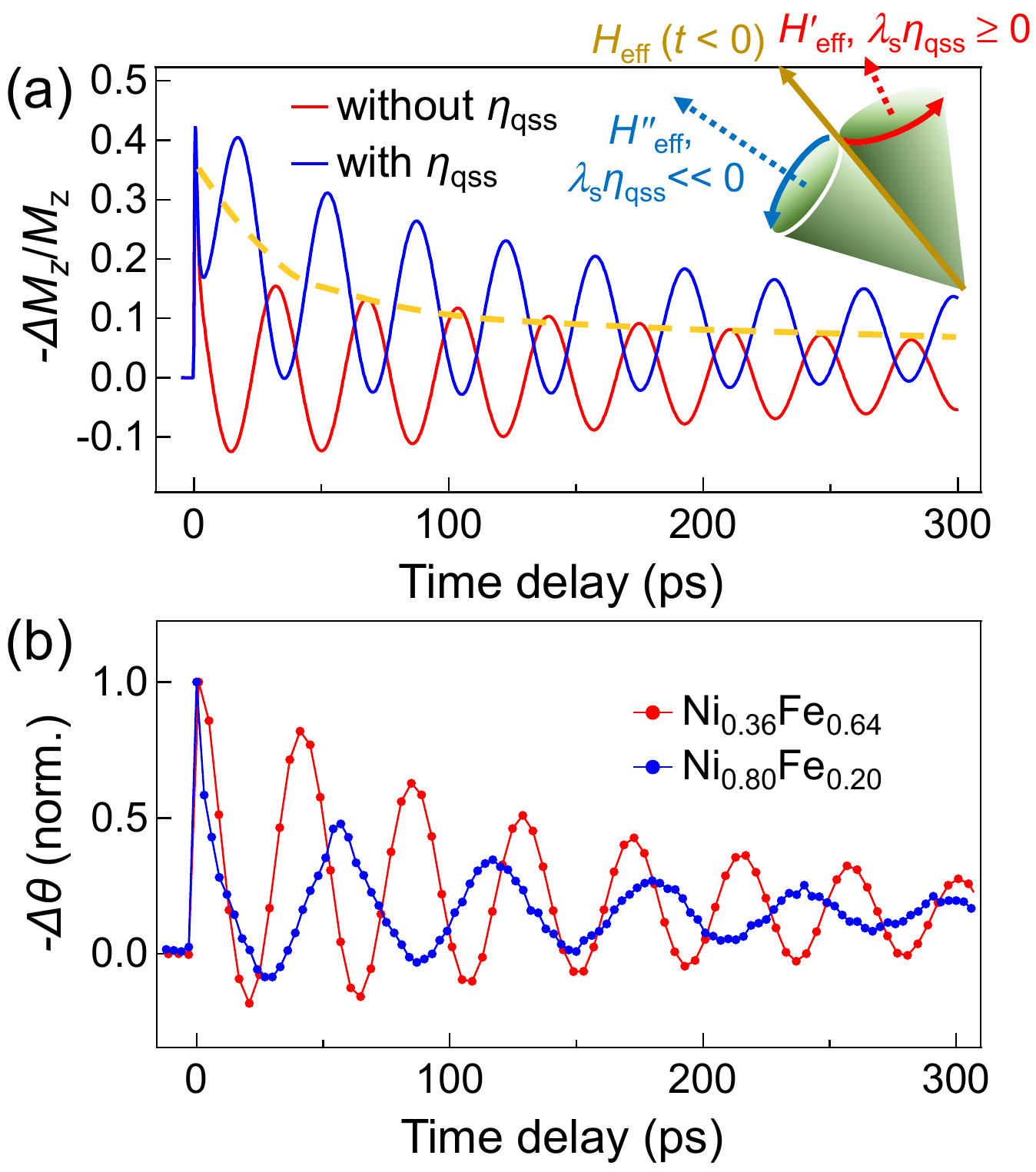}
\caption{QSS effect governing overall features of spin precession dynamics. (a) Calculated curves of spin dynamics for the absence (red) and presence (blue curve) of $\eta\textsubscript{qss}$ under $H$$\textsubscript{ext}$ = 0.54 T and $\varphi$ = 60 $^{\circ}$. The yellow dashed curve marks the monotonic-decaying apparent background level originated by a coherent rotation of $H$$\textsubscript{eff}(t)$ to the inplane direction, not by thermal magnons. Inset: schematic picture explaining the rotation direction of $H$$\textsubscript{eff}(t)$depending on either QSS (blue) or thermal (red) effect. (b) Experimental verification of the $\pi$-phase inversion determined at the initial stage of the precession for materials with a negligible magnetoelastic energy (Ni$_{0.36}$Fe$_{0.64}$ (red): $\beta$ $\sim$ 0, Ni$_{0.8}$Fe$_{0.2}$ (blue): $\lambda\textsubscript{s}$ $\sim$ 0).}\label{fig:fig4}
\end{figure}\\
\indent The pictorial description of the inset explains the $\pi$-phase inversion concisely. According to the conventional analysis, the sudden decrease of $M(t)$ after photo-excitation results in a rotation of $H\textsubscript{eff}(t)$ to the out of plane and starts spin precession around new axis of $H'\textsubscript{eff}(t)$. This aspect takes place when the thermal effect dominates the QSS effect for such conditions of $\lambda\textsubscript{s}$$\eta\textsubscript{qss}$ $\ge$ 0 ($\lambda\textsubscript{s}$ $\ge$ 0, $\beta$ $\ge$ 0 and $\lambda\textsubscript{s}$ $\le$ 0, $\beta$ $\le$ 0) and even $\lambda\textsubscript{s}$$\eta\textsubscript{qss}$ $\le$ 0 ($\lambda\textsubscript{s}$ $\le$ 0, $\beta$ $\ge$ 0) as long as the magnitude of $\lambda\textsubscript{s}$ is small. On the other hand, for $\lambda\textsubscript{s}$$\eta\textsubscript{qss}$ $\ll$ 0, the QSS effect prevails over the thermal effect and drives the rotation of $H\textsubscript{eff}(t)$ towards the inplane direction (blue arrow). The spins start the precession around the axis of $H''\textsubscript{eff}(t)$ of which magnitude is bigger than $H\textsubscript{eff}(t)$. This reorientation is determined at the initial stage of the precession, that is right after the demagnetization. As plotted in Fig. 4(b), we tested  Ni$\textsubscript{x}$Fe$\textsubscript{1-x}$(180)/Al$_{2}$O$_{3}$ films where it is predicted that the thermal effect dominates. As NiFe alloys for x $\sim$ 0.36 and 0.8 have negligible values of $\beta$ and $\lambda\textsubscript{s}$ respectively, it is expected that they have low magnetoelastic energy ($\propto$ $\lambda\textsubscript{s}$$\beta$$\Delta$$T$). Therefore, this case induces a rotation of $H\textsubscript{eff}(t)$ towards the out of plane direction $H'\textsubscript{eff}(t)$ after demagnetization, as the conventional analysis did.\\
\indent The second feature is about the pronounced background (yellow dashed line in Fig. 4(a)). This in general has been thought to consist of incoherent thermal magnons with large wavenumbers, which have the heat dissipation timescale and does not have a further meaning related to coherent motion of spins. However our results indicate that this should be interpreted as the contribution of a coherent rotation of $H\textsubscript{eff}(t)$ due to QSS effect. This is clearly verified from the fact that Co and Ni with high magnetoelastic energy have much larger exponential decaying offsets that those of NiFe alloys. These all features mentioned above including the frequency increase of the spin precession are reproduced through the model calculation incorporating only one concept of the QSS effect with full consistency to our experimental data.\\
\indent In summary, by decoupling the thermal and strain effect using ultrafast Sagnac interferometry and magneto-optical Kerr effect, we have proven that the quasi-static strain has a governing role over the thermal effect on the overall behavior of ultrafast spin dynamics. We demonstrate the ultrafast photo-excitation can make the effective magnetic field stronger with coherent counter-rotation to the inplane direction due to the quasi-static strain effect. This leads to a higher frequency of the spin precession and accounts for the $\pi$-phase inversion determined from the initial stage of the precession. Besides, the long-lived offset with the timescale of the quasi-static strain is interpreted by contribution of a coherent rotation of $H\textsubscript{eff}(t)$, not by incoherent thermal magnons. We emphasize that the quasi-static strain is involved universally in a wide family of materials and should be treated fundamentally. The strain of 0.1 $\sim$ 1 $\%$ generated by photo-excitations is high enough to induce the modification of the electronic band structure \cite{Akimov2006} even at ultrafast time scales \cite{Pudell2018}. Therefore it can be predicted that the quasi-static strain modifies the dielectric tensors bringing about new features such as derivative-like changes in differential magneto-optics leading to the inequivalence between its real and imaginary parts and differential reflectivity as well acting as involuntary extra gains even after thermal equilibrium timescale. To date, much effort has been dedicated to identifying the genuine magnetism and the origin of demagnetization \cite{Koopmans2000, Dornes2019, Bigot2004, Bigot2009, Battiato2010, Radu2009, Zhang2009, Koopmans2010, Siegrist2019}. Our demonstration that the quasi-static strain governs spin dynamics is the clear-cut result toward new perspective in the interpretation of ultrafast magneto-optics. We also hope that our work takes an important step by considering both strain and thermal effects for better understanding the physical mechanisms behind the ultrafast phenomena.
\begin{acknowledgments}
This work was supported by Basic Science Research Program through the National Research Foundation of Korea (NRF) funded by the Ministry of Education (2017R1A6A3A04011173 and 2020R1I1A1A01075040), and by Korea Research Foundation (NRF) (2018R1A2B3009569 and 2020R1A4A1019566), and by National Research Foundation of Korea (NRF) grant funded by the Korea government (MSIT) (2020R1A2C100613611), and by the Agence Nationale de la Recherche in France via the project EquipEx UNION No. ANR-10-EQPX-52. The authors gratefully thank G. Versini at IPCMS in CNRS for his effort to grow high quality samples.
\end{acknowledgments}

\bibliography{referencesPRL}

\end{document}